\documentclass[12pt,a4paper,twoside]{scrartcl}
\usepackage{graphicx}
\usepackage{amsmath}
\usepackage[figuresright]{rotating}
\usepackage{fancyhdr}
\usepackage{floatflt}
\newcommand{\TheAuthor}{}

\def\skiplinehalf{\medskip\\}

\def\supit#1{\raisebox{0.8ex}{\small\it #1}\hspace{0.05em}}  
\title{Muon Capture and Muon Lifetime}
\author{Peter Kammel\supit{a} \\
{\large \it representing the MuCap\cite{mucap} and MuLan\cite{mulan} collaboration}
\skiplinehalf
\supit{a}{\large University of Illinois at Urbana-Champaign, Urbana, IL 61801, USA}}
\date{}
\begin{document}
\maketitle
\begin{abstract}
We survey a new generation of precision muon lifetime experiments. The goal of the MuCap experiment
is a determination of the rate for muon capture on the free proton to 1 percent, from which the induced pseudoscalar 
form factor $g_P$ of the nucleon can be derived with 7 percent precision. A measurement of the
related $\mu$d capture process with similar precision would provide unique information on the axial current
in the two nucleon system, relevant for fundamental neutrino reactions on deuterium. The MuLan experiment
aims to measure the positive muon lifetime with 20 fold improved precision compared to present knowledge in order
to determine the Fermi Coupling Constant $G_F$ to better than 1 ppm.

\end{abstract}
\section{Overview}

A new generation of muon lifetime experiments is under preparation at the Paul Scherrer Institute
(see Table~\ref{table1}). The combination of novel experimental approaches, technological advances and
excellent beam quality promises a dramatic improvement in precision by typically an order of magnitude 
over earlier efforts. This paper surveys the scientific motivation and impact as well as the
strategy and status of these experiments.   
\begin{table}[h]
\vspace{-.2cm}
\begin{center}
\caption{Overview of experimental program of the MuCap and MuLan collaborations.}
\vspace{+.2cm}
\label{table1}
\renewcommand{\arraystretch}{1.2} 
\begin{tabular}{llll}
\hline
  Project      &   MuCap Experiment   &   $\mu$d Project   & MuLan Experiment\\
\hline 
  Physics      &   nucleon form factor, & EW reactions in 2-N               & fundamental constant \\
               &   chiral symmetry     & system, astrophysics       & of standard model     \\        
\hline 
Process   &  $\mu^- + p \to n + \nu_{\mu} $ &   $\mu^- + d \to n + n + \nu_{\mu} $ &  $\mu^+ \to e^+ + \nu_e +  \bar{\nu}_{\mu} $  \\ 
\hline
Observable & $\Lambda_S $ &  $\Lambda_D $ & $\tau_{\mu^+}$  \\
Precision Goal & $\le 1\% $& $\le 1\%$ &  $\le 1$ ppm \\ 
\hline
Physics Goal & $g_P \le 7\% $ & axial current, L$_{1A}$ &  $G_F \le 1$ ppm \\ 
\hline
\end{tabular} 
\end{center}
\end{table}
\section{Muon Capture on the Proton}
\subsection{Scientific Motivation}

Precision measurements of muon capture by the proton provide an excellent opportunity 
to probe the weak axial current of the nucleon and to test our understanding of chiral symmetry 
breaking in QCD. The importance of such experiments is underscored by the current
controversy between experiment and basic QCD predictions on the induced pseudoscalar 
coupling constant of the proton.

Ordinary muon capture (OMC) is a basic electroweak charged current reaction involving first-generation 
quarks and second-generation leptons
\begin{equation}
\mu^- + p \to n + \nu_{\mu}.
\label{eq:mucap}
\end{equation}
By virtue of Lorentz covariance and in the absence of second-class currents, 
the microscopic  electroweak structure of the nucleon can be parametrized by 
the four form factors $g_V$, $g_M$, $g_A$, and $g_P$
that determine the matrix elements of the charged vector and axial currents~\cite{ax02,fg03}
\begin{equation}
\bar{u}_n(p') \left[ g_V\gamma_\alpha+i\frac{g_M}{2 M_n}\sigma_{\alpha\beta}  q^\beta 
- (g_A\gamma_\alpha\gamma_5+\frac{g_P}{m_\mu}\gamma_5q_\alpha)\right] u_p(p).
\label{eq:current}
\end{equation}
The momentum transfer relevant for reaction~(\ref{eq:mucap}) is $q^2_0=(p'-p)^2=-0.88\ m^2_\mu$. 
The first three of these form factors are well determined by standard model symmetries and
 experimental data, leading to $g_V(q^2_0)=0.9755(5)$, $g_M(q^2_0)=3.582(3)$ 
and $g_A(q^2_0)=1.245(3)$. 
\begin{table}[htb]
\vspace{-.3cm}
\begin{center}
\caption{Recent calculations of $g_P \equiv g_P(q^2_0)$ and the capture rates $\Lambda_S$ and $\Lambda_T$ from the singlet and
triplet state of the $p\mu$ atom, respectively. 
Comparison of NLO and NNLO (next-to-next-to-leading order) calculations indicates good convergence of ChPT.}
\vspace{-.3cm}
\label{table2}
\renewcommand{\arraystretch}{1.2} 
\begin{tabular}{cclcccl}
\hline
Reference & Year     & & $g_P$ &   $\Lambda_S\;\;(\mathrm{s}^{-1})$ &  $\Lambda_T\;\;(\mathrm{s}^{-1})$ &Comment\\
\hline
\cite{ber94} & 1994  &       & $8.44(23)$ &&&chiral Ward identities \\
\cite{fea97} & 1997  &       & $8.21(9)$ & & & O(p$^3$) \\
\cite{gov00} & 2000  &       & $8.475(76)$ & $688.4 (3.8)$  &  $12.01(12)$\\
\cite{BHM00} & 2001  & NLO   & & 711 & 14.0 & small scale expansion\\
             &       & NNLO  & & 687.4 & 12.9 \\
\cite{AMK2000} & 2001& NLO   & & 722 & 12.2 & baryon ChPT\\
               &     & NNLO  & & 695 & 11.9 \\
\hline
\end{tabular} 
\end{center}
\vspace{-.5cm}
\end{table}

The induced pseudoscalar term is a direct consequence of the partial
conservation of the axial vector current. Its pion pole structure plus
the leading correction have been derived early on within current algebra~\cite{AD}.
During the last ten years $g_P$ has been studied with increasing
sophistication with baryon chiral perturbation theory (ChPT)~\cite{ax02,fg03}, i.e., within
a model-independent effective theory of QCD.  The theoretical results
are summarized in Table~\ref{table2} and indicate remarkably robust predictions for $g_P$ at 
the 2-3\% level, where the main 
uncertainty is related to the present knowledge of the pion--nucleon
coupling constant. Very recently a direct calculation
confirmed that the ChPT corrections at the two-loop order are small~\cite{ka03}.
Ref.~\cite{gov00} specifically addresses the sensitivity of muon capture experiments to
physics beyond the standard model. A similar body of recent theoretical work has
focused on the more intricate calculations of radiative muon capture in 
hydrogen (RMC)~\cite{ax02,fg03}.

\begin{figure}[htb]
\vspace{-1.1cm}
\begin{center}
\resizebox*{.7\textwidth}{.4\textheight}{\includegraphics{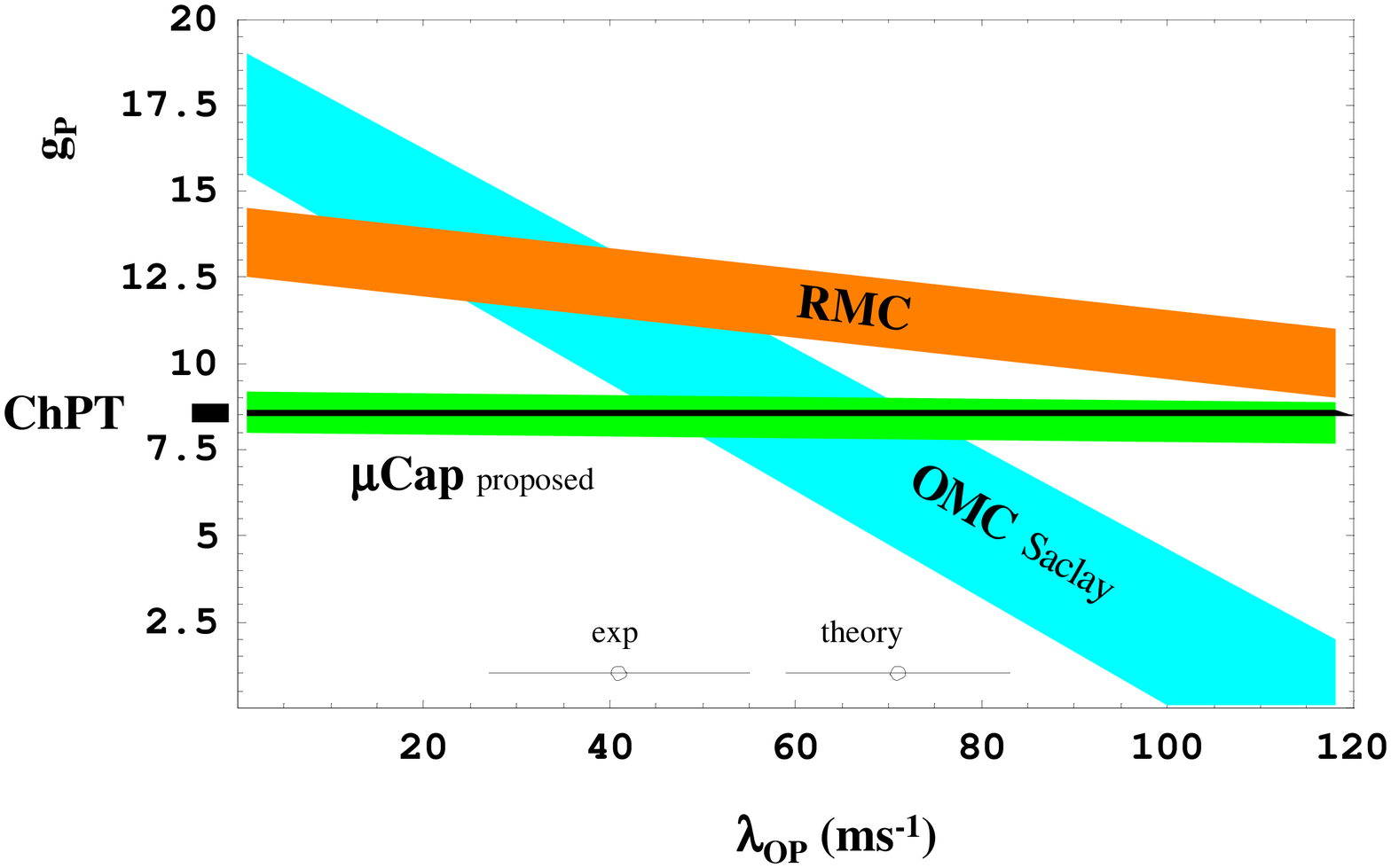}}
\vspace{-.7cm}
\caption{Current constraints on $g_P$ as function of $\lambda_{OP}$ from a recent 
updated analysis~\cite{fg03}. Experimental results from 
ordinary muon capture (OMC) \cite{OMC}, radiative muon capture (RMC) \cite{wri98}, 
and chiral perturbation theory (ChPT). The proposed MuCap experiment will be more precise
and less sensitive to $\lambda_{OP}$.}
\label{fig:gp}
\end{center}
\vspace{-.7cm}
\end{figure}

Despite more than 30 years of experimental effort in this field, existing OMC data only poorly 
constrain $g_P$. In addition to
the small rate $\Lambda_S$ 
and the all neutral final state of reaction~(\ref{eq:mucap}),
a main problem lies in the kinetics of muonic hydrogen. In order to achieve 
acceptable muon stop rates, most experiments were performed in
high density (i.e. liquid) targets, where muonic atoms quickly form $pp\mu$ molecules. The capture rates from these
states differ due to the strong spin dependence of the V-A interaction, so that
the uncertain transition rate $\lambda_{OP}$ between the molecular ortho and para state confuses the interpretation 
of observed capture rates. The recent pioneering measurement of the RMC process~\cite{wri98} is less sensitive
to $\lambda_{OP}$, but disagrees by 4.2 standard deviations from the accurate theory. Fig.~\ref{fig:gp} 
summarizes the present controversial situation~\cite{fg03}. In spite of
intense theoretical scrutiny, RMC could not be reconciled with theory, whereas even the best OMC
measurement cannot clarify this issue due to its dependence on muonic molecular physics.
The most accurate determination of $g_P$ in {\it nuclear} muon capture from   
a recent  experiment on $\mu^3\,$He capture \cite{mu3He} gives $g_p=8.53\pm1.54$ in good
agreement with theory. Note that the accuracy is limited by the theoretical extraction of $g_P$ from the 
three--nucleon system. A very recent calculation argues that this uncertainty in $g_P$ can be reduced to 6\% by
constraining the axial two-body current contributions from tritium beta decay~\cite{mar03}.

\subsection{The MuCap experiment}

\begin{figure}[htb]
\begin{center}
\resizebox*{0.45\textwidth}{0.26\textheight}{\includegraphics{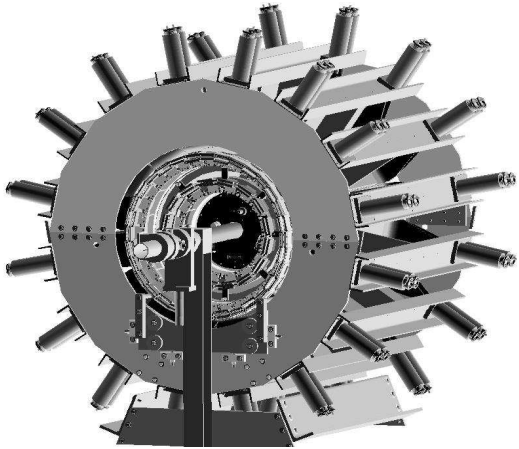}}
\caption{\label{mucap} MuCap detector showing scintillator barrel, cyclindrical wire chambers and vacuum pipe
supporting the hydrogen vessel.}
\end{center}
\vspace{-.8cm}
\end{figure}

The MuCap experiment~\cite{mucap} is based on a new method that avoids the above mentioned molecular and other
key uncertainties of earlier efforts, like the neutron detector calibration if the
outgoing neutron from reaction~(\ref{eq:mucap}) is observed.  
 The experiment is a muon lifetime measurement in ultra-pure
and deuterium-depleted hydrogen gas. The measured decay lifetime  $\tau _{\mu -}$
of the negative muon in hydrogen is shorter, compared with that of the positive muon  $\tau _{\mu +}$, 
because of the additional muon capture reaction.  
The rate $\Lambda _{S} =1/\tau _{\mu -} - 1/\tau _{\mu +} $
can be determined to 1\% if both the lifetime of the positive and negative muon are
measured with at least 10 ppm precision. Muons of both polarities will be stopped in an active target. Incoming muons
are tracked by wire chambers and are stopped in a specially developed time projection chamber (TPC) contained in a 
10-atm hydrogen pressure vessel. 
Two cylindrical wire chambers and a large scintillator hodoscope surround the TPC,  covering an effective solid angle
$\Omega/4\pi\sim 75\%$ (see Fig.~\ref{mucap}). 
This system will reconstruct the trajectories of the electrons from muon decay. 

Several unique features allow a dramatic improvement in precision:

{\bf \it Unambiguous interpretation.} As the target density is only 1\% of LH$_2$, $pp\mu$ formation, which
scales with density, is slowed. Muon capture takes place predominantly from the
singlet hyperfine state of the $p\mu$ atom and is nearly independent of  $\lambda_{OP}$ (see fig.~\ref{fig:gp}).

{\bf \it Clean muon stop definition.}
As the muon capture rate in higher-Z material can exceed $\Lambda_S$ by several orders of magnitude, it is essential
to eliminate muon stops and delayed diffusion to wall materials. By tracking the incident muons inside the TPC in
all three dimensions, the muon stop location is determined event by event. This eliminates wall stops and 
allows systematic off-line studies by selective cuts on the stopping distribution.
 
{\bf \it Gas purity control.}
Critical background reactions leading to charged recoils can
be monitored {\it in situ} with the TPC. This includes detection of
nuclear recoils in a $\mu + Z \to Z' + \nu $ capture, where a sensitivity to impurity levels of $\le 10^{-8}$
has been demonstrated. Also $p\mu \to d\mu$ transfer on isotopic deuterium impurities in the
target gas, which subsequently leads to significant $d\mu$ diffusion, can be directly seen in the data. 
In addition, a recirculating purification system is under development and gas analysis of higher-Z 
impurities and deuterium have been developed on the 0.01 and 1~ppm  sensitivity level, respectively.

{\bf \it High statistics.}
The detector system can operate with high muon stop rates up to 30~kHz using a custom designed, dead-time free
readout electronics for the TPC and the electron detectors. Pile-up effects, which have traditionally limited
the acceptable rate by causing high accidental background, are reduced by identifying muon electron pairs by their common vertex. 

{\bf \it $\mu SR$ rotation.}
The remnant polarization for positive muons introduces a position-dependent intensity variation of the decay positrons, 
which can affect the lifetime measurement if the detector is not perfectly uniform. A saddle coil magnet generates
a constant dipole field of $\sim$ 80 G in the target region, which will precess  the muon spin at 1 MHz.
Monte Carlo studies indicate that this sinusoidal part largely
decouples from the lifetime fit to the data.

\begin{figure}[htb] 
\vspace{-.2cm}
\begin{center}
\resizebox*{0.5\textwidth}{0.25\textheight}{\includegraphics{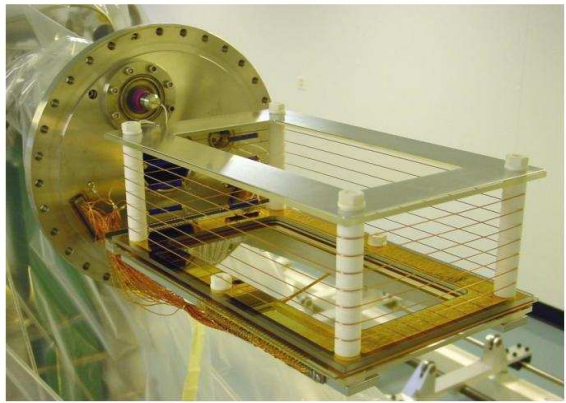}} 
\vspace{-0.3cm}
\caption{\label{tpc}  High-purity TPC. Drift volume height is 120 mm; area is 300$\times$150 mm$^2$. }
\end{center}
\vspace{-0.7cm}
\end{figure}
Several engineering runs with prototype TPCs have been performed at PSI to establish the feasibility of the new
method and to optimize physics and detector parameters~\cite{pmucap,kam00}.
Fig.~\ref{tpc} shows the new TPC during assembly in the hydrogen pressure vessel.
The design was optimized for the extremely stringent experimental purity requirements. 
Only UHV-proof materials were used, bakeable to 130$^\circ$ C. 
Chamber glass frames with metallic coatings were developed onto which the 
gold coated tungsten wires are soldered. The glass matches the small thermal expansion of tungsten. 
The drift voltage of ~30 kV produces a homogeneous electrical field of 2 kV/cm. 
In 10 bar hydrogen, this leads to drift velocities of $\sim$ 5 mm/$\mu$s. 
In the MWPC located at the bottom of the TPC, anode wires and cathode wires are read out giving 
two-dimensional coordinates. 
Operation at $\sim$ 6.5 kV high voltage results in a typical gas amplification of $10^4$.

The final hydrogen chamber system has been constructed and is presently being conditioned. The main part of the electron detector
was commissioned in fall of 2002 and was used for a high statistics $\mu^+$ lifetime measurement in a realistic set-up. This
included the precession  magnet, the full data acquisition chain and a selection of target materials to study $\mu SR$ effects.
For 2003, two runs are planned: a commissioning/integration run of the TPC and first physics data taking in fall. In 2004,
we foresee a high statistics MuCap run with a continuous beam or with the advanced Muon-On-Request beam~\cite{kam98}, which is
developed in the context of the MuLan experiment. 

\section{Muon Capture on the Deuteron}

\begin{figure}[hbt] 
\vspace{-1.2cm}  
\begin{center}
\resizebox*{1.\textwidth}{0.5\textheight}{\includegraphics{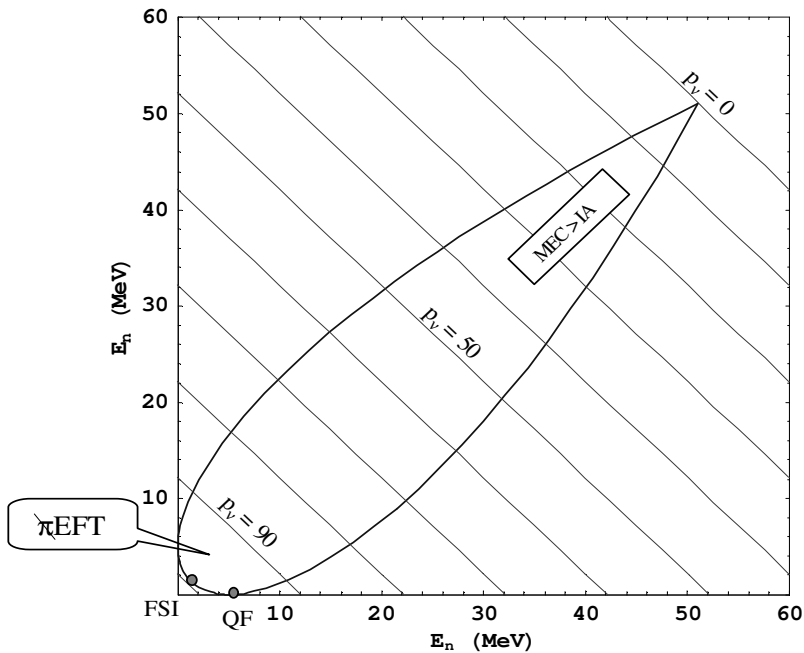}} 
\vspace{-1.6cm}
\caption{\label{dp}  The $\mu d$ capture Dalitz plot as function of neutron energy. Diagonal lines indicate
constant neutrino momentum $p_\nu$ (MeV/c). Interesting kinematic regions include: 
final state interaction (FSI) ; quasifree (QF) ; $p_\nu \ge $90 MeV/c, where pionless EFT applies; small  $p_\nu$, where
MECs dominate over impulse approximation.}
\end{center}
\vspace{-.7cm}
\end{figure}

Muon capture on the deuteron,
\begin{equation}
\mu^- + d \to n + n + \nu_{\mu},
\label{eq:mudcap}
\end{equation}

results from the same hadronic current as shown in Eq.~{\ref{eq:current}} and its rate $\Lambda_D$ from the doublet state of the 
$d\mu$ atoms shows a similar sensitivity to $g_P$
as $\Lambda_S$. There are additional interesting features. The final 3-body state covers a broad range
of momentum transfer to the 2-N system (see~Fig.~\ref{dp}).
The properties of the 2-N system enter, in particular the deuteron wave function and the neutron
scattering length $a_{nn}$ in the final state. Furthermore, two-body currents contribute, 
making process~(\ref{eq:mudcap}) uniquely suited to study the axial meson exchange
currents (MEC) in the 2-N system. 

In recent years the structure of the two-nucleon system and its response to electro-weak probes have received considerable 
attention. Progress has been made in complementary nuclear physics approaches, the calculation within the framework of
phenomenological Lagrangian models~\cite{pp}, the development of pionless effective theory~\cite{npi} 
of nucleon-nucleon interaction
and the hybrid MEEFT approach~\cite{meeft}, which incoporates both effective field theory (EFT) and Lagrangian models. These sophisticated 
calculations were
partially fueled by astrophysics interest, the importance of solar fusion $p+p \to d+e^+ +\nu$ as well as 
charged and neutral current $\nu+d$ scattering for the solar neutrino problem and the analysis of the SNO experiment. 
In particular, it
has been shown in model-independent EFT~\cite{nud}, that up to NNLO all neutrino deuteron break-up channels as well
as pp fusion are related by one isovector axial two-body current, parametrized by the counterterm L$_{1A}$. 
This stimulated intense effort~\cite{l1a} to calibrate all of these reactions by a single accurate measurement. Alas,
precision measurements of electroweak reactions in the 2-N system are extremely difficult, so that the 
most precise constraint of L$_{1A}$ is derived from the 3-N system~\cite{pp,l1a}. 

It was suggested by Kammel and Chen~\cite{kam00,chen}, that a determination of $\Lambda_D$ with $\sim$1\% precision
could provide the most accurate measurement of an electroweak reaction in the two-nucleon system and,
in particular, determine  L$_{1A}$. Such an experiment is under consideration by our collaboration.
In this context, one has to address at least two questions:
(a) Is reaction~(\ref{eq:mudcap}) soft enough so that it can be compared to
solar neutrino reactions? (b) Is a measurement at the required precision feasible?

A recent exploratory calculation~\cite{mud} within the framework of baryon ChPT constraining the axial 
MECs by tritium decay gives an affirmative answer to (a). $\Lambda_D$ can be calculated with ~1\% precision, while the 
contribution from the small $p_\nu$ region, where ChPT is questionable, is negligible. The L$_{1A}$ term contributes about 4\%
to the rate. On the other hand, the applicability of pionless EFT is limited to $p_\nu \ge $90 MeV/c and it would be
preferable to measure the reduced rate for this region~\cite{chen}.    

\begin{figure}[htb]
\vspace{-.5cm}
\begin{minipage}[t]{70mm}
{\resizebox*{1.1\textwidth}{.3\textheight}{\includegraphics{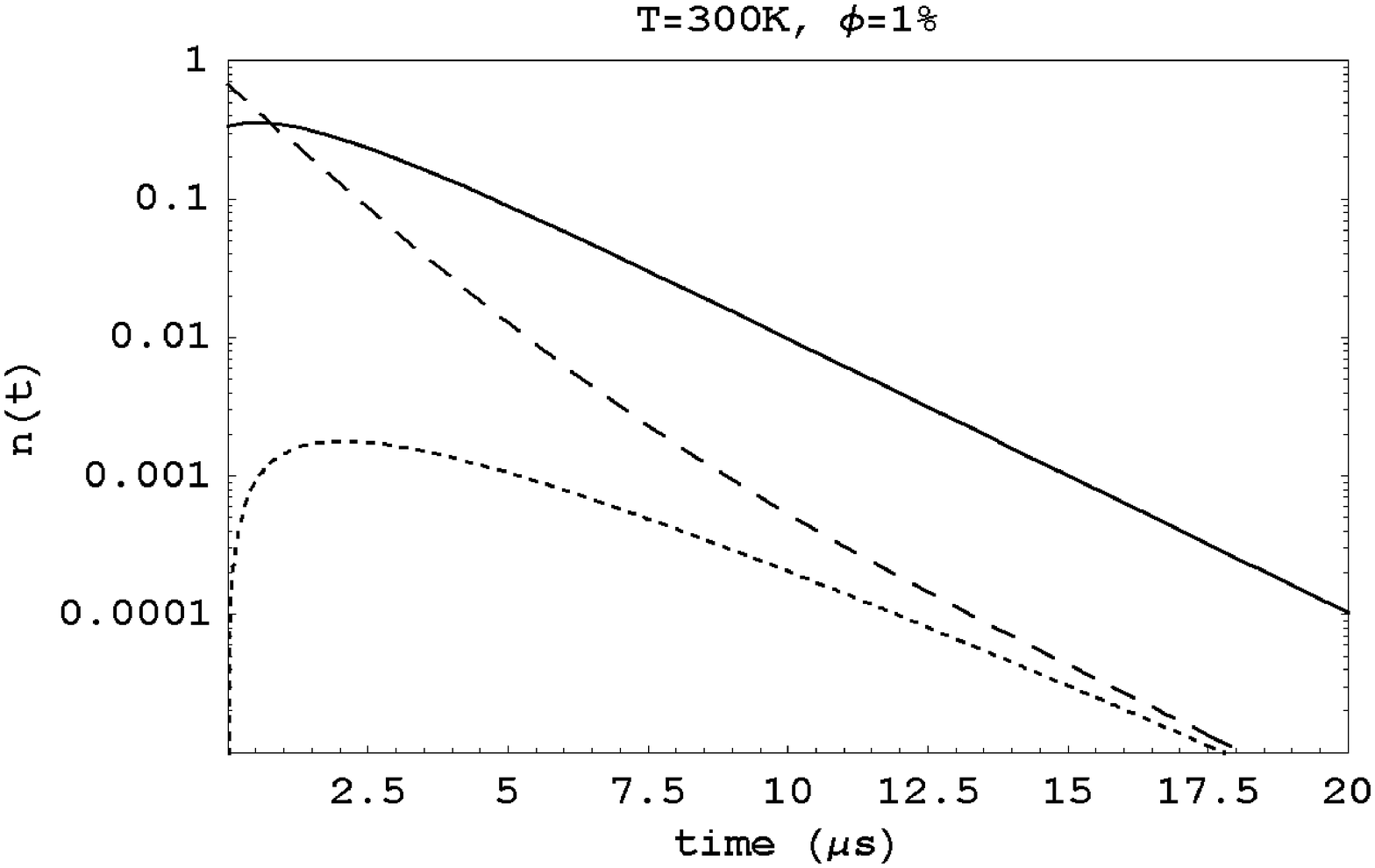}}}
\end{minipage}
\hspace{\fill}
\begin{minipage}[t]{70mm}
{\resizebox*{1.1\textwidth}{.3\textheight}{\includegraphics{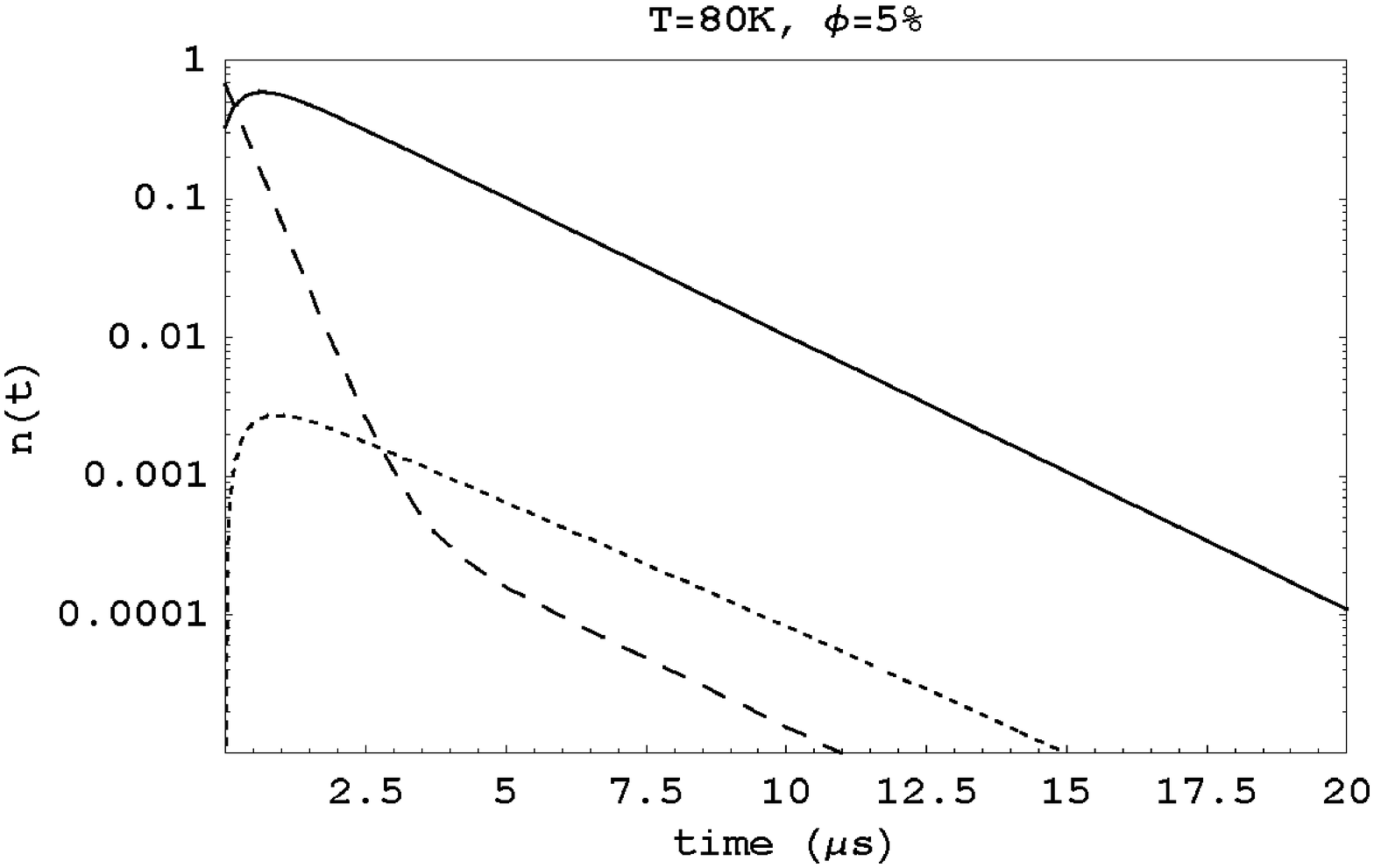}}}
\end{minipage}
\vspace{-.9cm}
\caption{\label{kinetics}  Time distributions of relevant states for different deuterium densities 
$\phi$ relative to LD$_2$ and temperatures. $d\mu(\uparrow \downarrow)$ (solid), $d\mu(\uparrow \uparrow)$ (dashed), 
$^3He \mu$ (dotted).}
\vspace{-.2cm}
\end{figure}
As regards (b) current experimental results are of 6-10\% precision and
only marginally agree with each other~\cite{mudexp}. However, the MuCap strategy should allow significant
improvement. Moreover - if necessary - the reduced rate for the kinematic region $p_\nu \ge $90 MeV/c can be determined as the 
difference between a precise measurement of the total rate $\Lambda_D$ (via the lifetime method) and a measurement of the 
Dalitz plot of neutrons with E$_n \ge $10 MeV (with neutron detectors). The latter measurement needs only limited 
precision of 5-10\%, as the high-energy part contains only a small fraction of the total intensity.       
The main challenge is, in a different guise as for $\mu+p$ capture, the muon induced kinetics. 
The hyperfine quenching of the upper $d\mu(\uparrow \uparrow)$ quartet to the $d\mu(\uparrow \downarrow)$ doublet state is slow. 
In addition, $^3\,$He$\mu$ atoms are
formed after $dd\mu$ formation and fusion, where muon capture is larger than in deuterium.
The $d\mu$ system has been intensively studied as the prototype for resonant muon-catalyzed 
fusion~\cite{mcf}.  For a clean interpretation
the target conditions should be chosen such that the $d\mu(\uparrow \downarrow)$ state dominates and the 
population of states can be verified {\em in-situ} by the observation of muon-catalyzed fusion reactions.
A  preliminary optimization (Fig.~\ref{kinetics}) indicates promising conditions at 
$\phi=5\%$ and T=80 K. 

\section{Muon Lifetime}
\subsection{Scientific Motivation}

The Fermi Coupling Constant $G_F$ is a fundamental
constant of nature. In particular, together with $ \alpha = 1/137.03599976(50)\; (0.0037\,{\rm ppm})$ and
$M_{\mathrm{Z}} =91.1876(21){\rm GeV}\; (23\,{\rm ppm})$,  $G_F$ 
defines the gauge couplings of the electroweak sector of the standard model. The most precise determination of
$G_F$ comes from the measurement of the muon lifetime $\tau_\mu$~\cite{stu,mar99}
\begin{equation}
\frac{1}{\tau_\mu}=\frac{G_F^2 m_\mu^5}{192\pi^3}(1+\Delta q).
\label{eq:QEDcorr}
\end{equation} 
Here
$\Delta q$ encapsulates the higher order QED and QCD corrections calculated in the Fermi theory.
The remaining electroweak corrections are contained in the quantity~\cite{sir80}
 $\Delta r$ defined by
\begin{equation}
\frac{G_F}{\sqrt{2}}=\frac{g^2}{8 M_{\mathrm{W}}^2}\left(1+\Delta r\right)
\label{eq:GFdef}
\end{equation} where $g$ and $M_{\mathrm{W}}$ are the $SU(2)_L$
coupling constant and the W boson mass, respectively. Interesting
quantum loop effects are absorbed in this quantity, including a remarkable
sensitivity to the top quark mass and to the Higgs mass as well
as potential new physics~\cite{mar99}. At the moment,
$G_F$ is not the limiting factor in exploiting these loop contributions,
as the precision of other electroweak observables still has to be improved considerably.
Recent 2-loop QED calculations~\cite{stu,sir99}
led to a revised value and error of $G_F = 1.16637(1)\times10^{-5}\,{\rm GeV^{-2}} (9\,{\rm ppm})$. 
The hitherto dominant theoretical uncertainty in Eq.~\ref{eq:QEDcorr} was reduced to a negligible level.  
Thus an extraction of $G_F$ from experiment  down to the level of 0.5 ppm is becoming feasible 
before encountering  theoretical limitations. This fact together with the spectacular
precision achieved for other electroweak parameters, most notably the Z mass, has stimulated a new generation 
of $\mu^+$ lifetime experiments at PSI~\cite{mulan,fast} and Rutherford Appleton Laboratory~\cite{ral}.

The extraction of $G_F$ from Eq.~\ref{eq:QEDcorr} involves the following experimental contributions
\begin{equation}
\begin{array}{lllll}
\frac{\delta G_F}{G_F} &=&
                       \sqrt{(\frac{5}{2}\frac{\delta m_\mu}{m_\mu})^2
                       +(\frac{1}{2}\frac{\delta\tau_\mu}{\tau_\mu})^2 
                       \;[+(4\frac{m_{\nu_\mu}^2}{m_\mu^2})^2]}
                       &=& \sqrt{0.38^2+9^2\; [ +10^2] \;} ppm \\
\end{array}
\end{equation}
The last contribution is based on the current upper bound of $m_{\nu_\mu}\le170$\,keV from direct experiments.
However, given the present empirical information on neutrino masses from $\nu$-oscillation and cosmic microwave 
background measurements, such a large $m_{\nu_\mu}$ appears unrealistic. Thus the main uncertainty in
$G_F$ is due to $\tau_\mu$.

\subsection{The MuLan experiment}

The MuLan collaboration~\cite{mulan} plans to reduce the uncertainty of the present
experimental value of 2197.03$\pm$ 0.04 ns (18\,ppm) to 1 ppm.
Both the statistics and systematics must be dramatically improved, relying on
the following experimental concept.
\begin{figure}[h]
\begin{center}
{\resizebox*{0.7\textwidth}{0.18\textheight}{\includegraphics{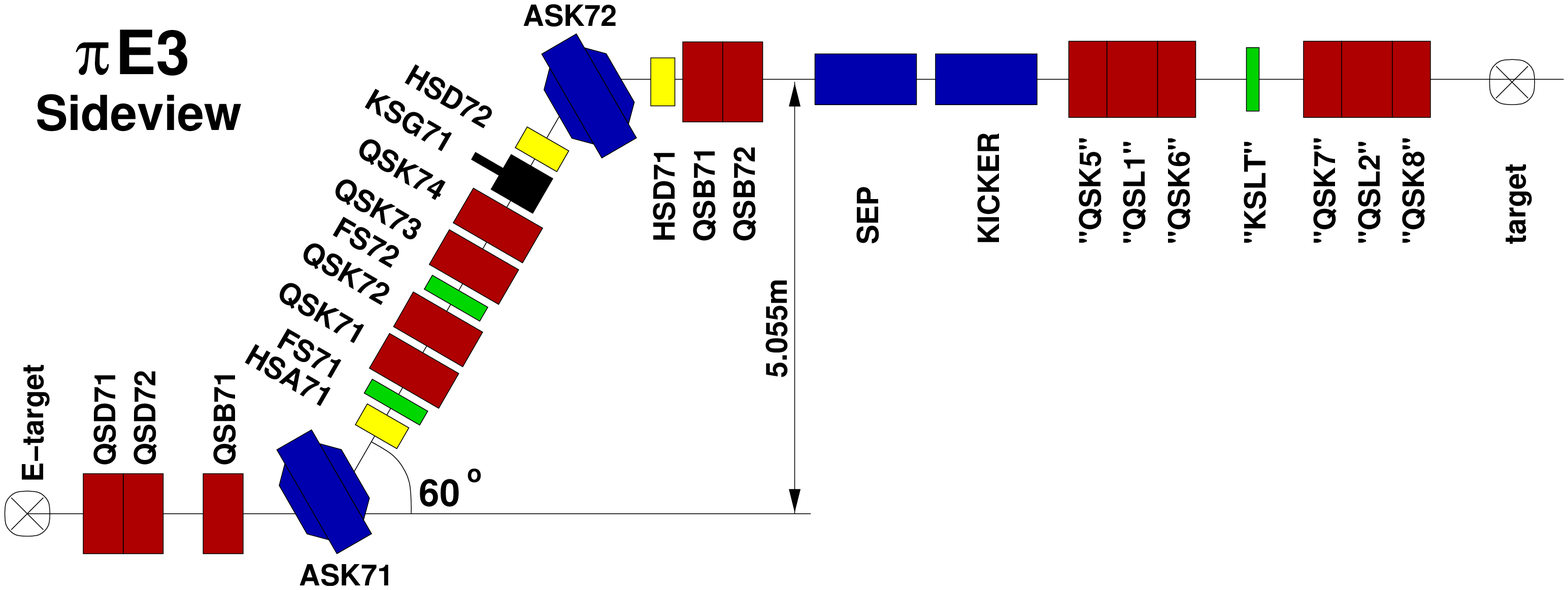}}}
{\resizebox*{0.7\textwidth}{.2\textheight}{\includegraphics*[angle=-90]{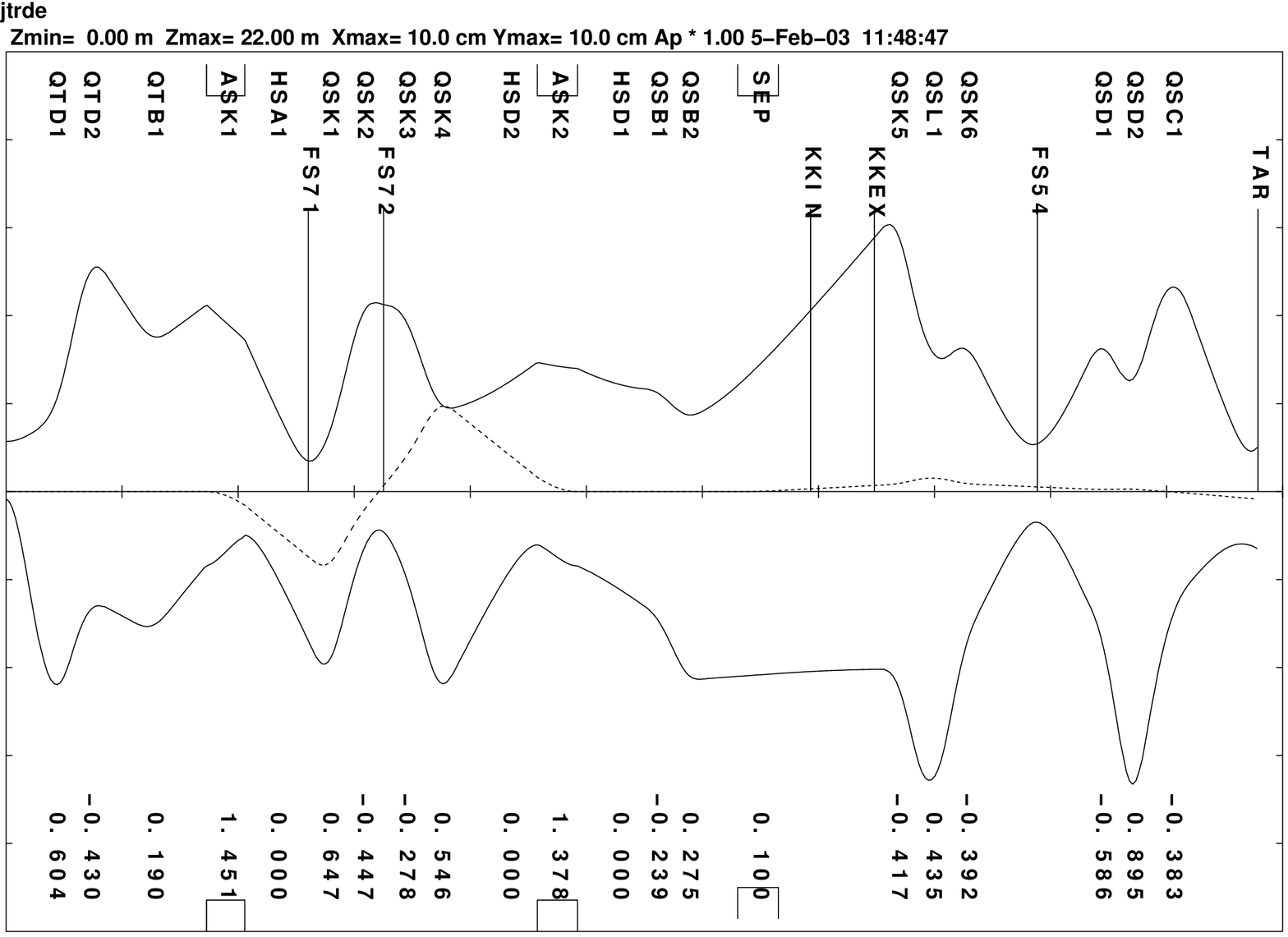}}}
\caption{\label{pie3}  New $\pi E3$ beamline and calculated beam envelopes used in chopper tests.}
\end{center}
\vspace{-.7cm}
\end{figure}

{\em Statistics of 10$^{12}$ events}. This statistics cannot be achieved with
the conventional ``one-muon-at-a-time'' method. Instead, muons from the
continuous PSI beam will be electrostatically chopped~\cite{kam98} at $\sim$ 50 kHz in the secondary 
surface muon beam line of the high intensity $\pi$E3 area. This allows
to accumulate pulses with several muons 
and then have beam free interval of $\sim 20 \,\mu s$ for measurement.  

{\em Pile-up suppression}. 
Decay positrons are registered in a soccer ball shaped
arrangement of 170 tile elements consisting of 3 mm
thick fast plastic scintillator pairs, which covers $\sim 75\%$ of $4\pi$. 
The average inner diameter of the ball is $\sim$ 80 cm. 
The analog signals from the photomultipliers coupled to the scintillators will be digitized
by 8-bit 500 MHz waveform digitizers. High segmentation, excellent pulse
pair resolution and energy information are crucial to identify and suppress pile-up of two electrons 
hitting a tile simultaneously, which would systematically distort
the observed time spectrum.

{\em Detector stability}.
 Target and detector material are minimized
and low thresholds are used, so that the majority of Michel 
electrons hitting a tile are detected and the effect of time-dependent
gain/threshold changes is reduced. 

{\em $\mu SR$ effect}. As discussed for MuCap, residual $\mu^+$ polarization and
subsequent precession and depolarization can lead to systematic distortions
of the decay spectra. This effect is suppressed in several steps: by the highly symmetric
detector geometry, the choice of depolarizing target material and the use of a  
transverse magnetic field to reduce the overall ensemble polarization per 
pulse and to monitor the muon spin asymmetry.

The effect of muonium
formation to the lifetime has been found to be negligible~\cite{muonium}.

A recent milestone for MuLan was the development of a new kickable tune of
the $\pi E3$ channel, based on extensive phase space simulations and measurements.
The tune minimizes the angular divergence in the kicker plane, while simultaneously
optimizing the phase space acceptance to obtain the required muon flux. 
The measured extinction factor of $\le 10^{-3}$ at a beam intensity of $\sim$15 MHz
lies comfortably within the proposed design specifications. The new chopper is presently
being built at TRIUMF and PSI. 
Two MuLan runs are scheduled for 2004, the first for commissioning the chopper and beamline and
the second for commissioning the newly built detector, electronics and high rate data acquisition.
The new chopper supports different time structures, providing an exciting new facility for MuLan, 
MuCap and other fundamental muon experiments.


\begin{thebibliography}{9}
\footnotesize
\bibitem{mucap}
V.A. Andreev, A.A. Fetisov, V.A. Ganzha, V.I. Jatsoura, A.G. Krivshich, E.M. Maev, O.E. Maev, 
G.E. Petrov, S. Sadetsky, G.N. Schapkin, G.G. Semenchuk, M. Soroka, A.A. Vorobyov
(Petersburg Nuclear Physics Institute (PNPI), Gatchina 188350, Russia),
P.U. Dick, A. Dijksman, J. Egger, D. Fahrni, M. Hildebrandt, A. Hofer, 
L. Meier, C. Petitjean, R. Schmidt
(Paul Scherrer Institute, PSI, CH-5232 Villigen, Switzerland), 
T.I. Banks, T.A. Case, K.M. Crowe, S.J. Freedman, B. Lauss
(University of California Berkeley, UCB and LBNL, Berkeley, CA 94720, USA),
K.D. Chitwood, S. Clayton, P. Debevec, F. E. Gray, D. W. Hertzog, P. Kammel, 
B. Kiburg, C. J. G. Onderwater, C. Ozben, C. C. Polly, A. Sharp
(University of Illinois at Urbana-Champaign, Urbana, IL 61801, USA), 
L. Bonnet, J. Deutsch, J. Govaerts, D. Michotte, R. Prieels
(Universit\'{e} Catholique de Louvain, B-1348 Louvain-La-Neuve, Belgium), 
R.M. Carey, J. Paley
(Boston University, Boston, MA 02215, USA),
T. Gorringe, M. Ojha, P. Zolnierzcuk
(University of Kentucky, Lexington, KY 40506, USA), 
F.J. Hartmann 
(Technische Universit\"{a}t M\"{u}nchen, D-85747 Garching, Germany), 
``High precision measurement of the singlet
$\mu p$ capture rate in $H_2$ gas'', PSI proposal R-97-05, http://www.npl.uiuc.edu/exp/mucapture
%
\bibitem{mulan}
R. Carey, A. Gafarov, I. Logachenko, K. Lynch, J. Miller, L. Roberts
(Boston University, Boston, MA 02215, USA), 
D. Chitwood, S. Clayton, P. Debevec, F. Gray, D. Hertzog, P. Kammel, B. Kiburg, G. Onderwater, C. Ozben,
C. Polly, A. Sharp, S. Williamson
(University of Illinois at Urbana-Champaign, Urbana, IL 61801, USA),
M. Deka, T. Gorringe, M. Ojha
(University of Kentucky, Lexington, KY 40506, USA), 
K. Giovanetti (James Madison University, Harrisonburg,  VA 22807, USA), 
K. Crowe, B. Lauss(University of California Berkeley, UCB and LBNL, Berkeley, CA 94720, USA),
``A Precision Measurement of the Positive Muon Lifetime Using a
Pulsed Muon Beam and the $\mu$Lan Detector'', PSI proposal R-99-07, http://www.npl.uiuc.edu/exp/mulan
%
\bibitem{AD} S.L.~Adler, Y.~Dothan, 
             Phys. Rev. {\bf 151} (1966) 1267; 
             L.~Wolfenstein, in {\it High-Energy Physics and Nuclear Structure}, 
             Plenum, NY, 1970, p.~661.
%
\bibitem{ax02}
V.~Bernard, L.~Elouadrhiri and U.~G.~Meissner,
J.\ Phys.\ G {\bf 28} (2002) R1.
%
\bibitem{fg03}
T.~Gorringe and H.~W.~Fearing,
arXiv:nucl-th/0206039, submitted to Rev. Mod. Phys.
%
\bibitem{ber94}
V.~Bernard, N.~Kaiser and U.G.~Meissner, {\it Phys. Rev. D}{\bf 50}, (1994) 6899.
%
\bibitem{fea97}
H.W.~Fearing {\it et al.\/}, {\it Phys. Rev.} {\bf D 56} (1997) 1783.
%
\bibitem{gov00} J.~Govaerts and J.-L.~Lucio-Martinez, Nucl. Phys.
        {\bf A678} (2000) 110.
%
\bibitem{BHM00}
V.~Bernard, T.R.~Hemmert and U.G.~Meissner, Nucl. Phys.  {\bf A686} (2001) 290. 
%
\bibitem{AMK2000} S.~Ando, F.~Myrer and K.~Kubodera, 
                  Phys. Rev. {\bf C 63} (2001) 015203. 
%
\bibitem{ka03}
N.~Kaiser,
Phys.\ Rev.\ C {\bf 67} (2003)  027002.
%
\bibitem{OMC} G.~Bardin {\it et al.,} 
              Phys. Lett. {\bf B 104} (1981) 320.
%
\bibitem{wri98}
D.H. Wright {\it et al.}, Phys. Rev. {\bf C 57} (1998) 373.
%
\bibitem{mu3He} P.~Ackerbauer {\it et al.}, 
                Phys. Lett. {\bf B417} (1998) 224. 
%
%
\bibitem{mar03}
L.~E.~Marcucci  {\it et al.\/} ,
Phys.\ Rev.\ C {\bf 66} (2002) 054003.
%
\bibitem{pmucap}
E.~M.~Maev {\it et al.}, Nucl.\ Instrum.\ Meth.\ A {\bf 478} (2002)  158; Hyperfine Interactions 138 (2001) 451;
P. Kammel {\it et al.},  Nucl. Phys. {\bf A 663 \& 664}, 911c (2000) ; 
A.A. Vorobyov {\it et al.}, Hyp. Int. {\bf 119} (1999) 13;
P. Kammel AIP Conf. Proc. 435 (1998) 419.
%
\bibitem{kam00}
P. Kammel {\it et al.}, Hyperfine Interactions 138 (2001) 435.
%
\bibitem{kam98}
P. Kammel {\it et al.\/},
High intensity muon/pion beam with time structure at PSI,
PSI letter of intent, R-98-04.0, 1998, 
Hyperfine Interactions {\bf 119} (1999) 323.
%
\bibitem{pp}
R. Schiavilla {\it et al.\/}, Phys Rev. {\bf C 58} (1998) 1263.
%
\bibitem{npi}
J.~W.~Chen, G.~Rupak and M.~J.~Savage,
Nucl.\ Phys.\ A {\bf 653} (1999) 386.
%
\bibitem{meeft}
T.~S.~Park {\it et al.}, arXiv:nucl-th/0208055.
\bibitem{nud}
M.~Butler, J.~W.~Chen and X.~Kong,
Phys.\ Rev.\ C {\bf 63}  (2001) 035501.
%
\bibitem{l1a}
M.~Butler, J.~W.~Chen and P.~Vogel,
Phys.\ Lett.\ B {\bf 549} (2002) 26 ;
J.~W.~Chen, K.~M.~Heeger and R.~G.~Robertson,
Phys.\ Rev.\ C {\bf 67}  (2003) 025801. 
%
\bibitem{mud}
S.~Ando, T.~S.~Park, K.~Kubodera and F.~Myhrer,
Phys.\ Lett.\ B {\bf 533} (2002) 25.
%
\bibitem{chen}
W.J. Chen, work in progress.
%
\bibitem{mudexp}
G. Bardin {\it et al.}, Nucl. Phys. A 453 (1986) 591;
M. Cargnelli {\it et al.},  
Proceedings of the XXIII Yamada Conf. on Nuclear Weak Processes and
Nuclear Structure, Osaka (1989), eds. M. Morita, E. Ejiri, H.Ohtsubo
and T. Sato (World Scientific), p.115
%
\bibitem{mcf}
W.H. Breunlich, P. Kammel, J.S. Cohen and M. Leon, 
Annu. Rev. Nucl. Part. Sci. 39  (1989) 311;
P. Kammel {\it et al.}, Phys. Rev. {\bf A 28}  (1983) 2611; 
J. Zmeskal {\it et al.}, Phys. Rev. A 42,(1990) 1165; A. Scrinzi {\it et al.}, 
Phys. Rev. {\bf A 47} (1993) 4691;
N. I. Voropaev {\it et al.}, Hyperfine Interactions {\bf 138} (2001) 331. 



%
\bibitem{stu}
T.~van Ritbergen and R.G.~Stuart,
Phys.\ Lett.\ {\bf B437} (1998) 201;
P.~Malde and R.~G.~Stuart,
Nucl.\ Phys.\ B {\bf 552} (1999) 41;
T.~van Ritbergen and R.~G.~Stuart,
Nucl.\ Phys.\ B {\bf 564} (2000) 343. 
%
\bibitem{mar99}
W.~J.~Marciano,
Phys.\ Rev.\ D {\bf 60} (1999) 093006;
J.\ Phys.\ G {\bf 29} (2003) 23.
%
\bibitem{sir80}
A. Sirlin, {\sl Phys.\ Rev.}\ {\bf D 22} (1980) 971.
%
\bibitem{sir99}
A.~Ferroglia, G.~Ossola and A.~Sirlin,
Nucl.\ Phys.\ B {\bf 560} (1999) 23. 
%
\bibitem{fast}
F.R. Cavallo  {\it et al.\/},  Precision measurement
of the $\mu^+$ lifetime ($G_F$) with the FAST detector, PSI proposal R-99-06 (1999). 
%
\bibitem{ral}
S. N. Nakamura  {\it et al.\/}, Hyperfine Interactions 138 (2001) 445.
%
\bibitem{muonium}
A.~Czarnecki, G.~P.~Lepage and W.~J.~Marciano,
Phys.\ Rev.\ D {\bf 61} (2000) 073001.


\end{thebibliography}
\end{document}